\documentstyle[epsf,12pt,fullpage]{article}

\def\bmat	{\left( \begin{array}}
\def\emat	{\end{array} \right)}
\def\beqn	{\begin{eqnarray*}}
\def\eeqn	{\end{eqnarray*}}
\def\beqa	{\begin{eqnarray}}
\def\eeqa	{\end{eqnarray}}
\def\bitm	{\begin{itemize}}
\def\eitm	{\end{itemize}}
\def\beq	{\begin{equation}}
\def\eeq	{\end{equation}}

\def\f		{\frac}
\def\rarw	{\rightarrow}

\def\ran 	{\rangle}
\def\lan 	{\langle}

\newcommand{\av}[1]     {\langle #1 \rangle}

\def\a		{\alpha}
\def\b		{\beta}

\def\e		{\epsilon}
\def\g		{\gamma}
\def\d		{\delta}

\def\m		{\mu}
\def\n		{\nu}
\def\s		{\sigma}

\def\D		{\Delta}

\def\bk		{{\bf k}}
\def\bl		{{\bf l}}
\def\br		{{{\bf r}}}

\def\br         {{\bf r}}

\def\bA		{{\bf A}}

\def\bQ		{{\bf Q}}

\def\tr		{\mbox{tr~}}

\def\cis	{C_{i\sigma}^{}}
\def\cisd	{C_{i\sigma}^\dagger}

\def\cjs	{C_{j\sigma}^{}}

\def\cmns	{C_{(j_x, j_y)\s}}

\def\cmnsd	{\cmns^{\dagger}}

\def\cmpns	{C_{(j_x\!+1, j_y)\s}}

\def\cmpnsd	{\cmpns^{\dagger}}

\def\cmmns      {C_{(j_x\!-1,j_y)\s}}

\def\cmmnsd      {\cmmns^{\dagger}}

\def\cmnps	{C_{(j_x, j_y+1)\s}}

\def\cmnpsd	{\cmnps^{\dagger}}

\def\cmnms	{C_{(j_x, j_y-1)\s}}

\def\cmnmsd	{\cmnms^{\dagger}}
\def\phc	{2\pi \f{p}{q}j_x i}
\def\phn	{(2\pi \f{p}{q}j_x +k_ya)i}

\def\niu    {n_{i\uparrow}}
\def\nid    {n_{i\downarrow}}

\def\tij	{t_{ij}^{}}

\def\sumi	{\sum_{i}}

\def\summns	{\sum_{(j_x, j_y)\s}}

\def\sumijs	{\sum_{\langle ij\rangle\sigma}}
\def\sumis  {\sum_{i\sigma}}

\def\ketpsi		{|\psi\ran}
\def\ketmn		{|(j_x, j_y)\ran}

\def\mzpsi	{\lan (j_x,\: 0)|\psi\ran}
\def\mopsi	{\lan (j_x,\: 1)|\psi\ran}

\def\mnpsi	{\lan (j_x,\: j_y)|\psi\ran}

\def\mpnpsi	{\lan (j_x\!\!+\!\!1,\: j_y)|\psi\ran}
\def\mnppsi	{\lan (j_x,\: j_y\!\!+\!\!1)|\psi\ran}
\def\mmnpsi	{\lan (j_x\!\!-\!\!1,\: j_y)|\psi\ran}
\def\mnmpsi	{\lan (j_x,\: j_y\!\!-\!\!1)|\psi\ran}

\begin{document}
\title{
Effects of Electron Correlations on Hofstadter Spectrum
}
\author{Hyeonjin Doh and Sung-Ho Suck Salk\\
Department of Physics, Pohang University of Science and Technology,\\
Pohang 790-784, Korea}
\maketitle
\begin{abstract}
By allowing interactions between electrons, a new Harper's equation is derived to examine the effects of electron correlations 
on the Hofstadter energy spectra.
It is shown that the structure of the Hofstadter butterfly for the system of 
correlated electrons is modified only in the band gaps and the band widths, 
but not in the characteristics 
of self-similarity and the Cantor set. 
\end{abstract}
\newpage
\section{Introduction}
\hspace*{0.5cm}Numerous studies have been made to study two dimensional 
quantum systems of electrons in a magnetic field, including the Hofstadter
energy spectrum [1] and the quantum Hall effects [2]. 
Hofstadter studied the energy spectrum for non-interacting electrons on a two dimensional square lattice
in a perpendicular uniform magnetic field, namely the energy spectrum
vs. magnetic flux per plaquette known as the Hofstadter butterfly.
The energy spectrum was found to critically
depend on the ratio $p/q$ ($p$ and $q$ are positive integers) of the magnetic 
flux per plaquette to the flux quantum.
If $p/q$ is a rational
 number, 
each energy band is split into $q$ sub-bands by the magnetic field.
The Hofstadter butterfly 
displays a recursive structure over rational (number) fields and a Cantor set at 
any irrational number of $p/q$ [1].

The gap structure predicted by the Hofstadter butterfly is known to play an important role on the magnetoresistance of the laterally modulated two-dimensional electron systems in GaAs-AlGaAs heterostructure [3]. 
Such gap structure can be sensitive to the strength of interactions between electrons.
Lately, based on the Hartree approximation Gudmundsson and Gerhardts [4]  
studied the effects of interacting electrons on the Hofstadter butterfly.  
In the present study we derive a new Harper's equation 
which will allow us to study the Hofstadter energy spectrum as a function of Coulomb repulsion by properly taking into account electron interactions. 
As an illustration we pay a special attention to the interesting 
system of antiferromagnetically correlated electrons on a square lattice.

\section{Hofstadter Spectrum of Interacting Electrons}

\hspace*{0.5cm}Earlier Hasegawa et al.[5] studied the Harper's equation [6] and the energy spectrum
 only for spinless non-interacting electrons.
In the present study we derive a new Harper's equation in order to study the Hofstadter butterfly, that is, the energy spectrum vs.
magnetic flux as a function of interaction  strength (Coulomb repulsion $U$) between electrons.

In order to treat interacting electrons in 
a two dimensional square lattice in a perpendicular uniform magnetic field,
we consider the one-band Hubbard model Hamiltonian,
\beq
H=-\sumijs \tij \cisd \cjs
 + U \sumi \niu \nid - \m\sumis \cisd\cis,
\label{flux_mean}
\eeq
with
\beq
t_{ij} = -t\exp\left(-i\f{2\pi}{\phi_0}\int_j^i \bA\cdot d\bl\right).
\eeq
Here $C^\dag_{i\s}(C_{i\s})$ is the 
creation (annihilation) operator of an electron with spin $\s$ at site $i=(i_x,~i_y)$.
$\lan ij \ran$ denotes only the nearest neighbor hopping with lattice spacing $a$.
$\phi_0$ is the unit flux.
The choice of the Landau gauge, $\bA = B(0,j_x a,0)$ for a uniform 
magnetic field $B$ leads to
the hoping integrals, 
$t_{ij}=t_{(j_x\!\pm\! 1, j_y)(j_x, j_y)} = t$ for the x-direction of hopping and
$t_{ij}=t_{(j_x, j_y\!\pm\! 1)(j_x, j_y)} = t\exp\left(\mp i2\pi j_x\f{\phi}{\phi_0}\right)= t\exp 
\left(\mp 2\pi j_x\f{p}{q}\right)$ for the y-direction of hopping,
where $\phi\equiv B a^2$. 
$\f{\phi}{\phi_0}=\f{p}{q}$ with $p$ and $q$ integers is the 
number of flux quanta per plaquette which is equivalent to the gain of phase by a particle as a result of hopping round the closed path of a plaquette. 

We now introduce the staggered magnetization at site $i$,
$$
m(i)=e^{i\bQ\cdot\br}\sum_{\s}\s\av{\cisd\cis}
$$
with $\bQ \equiv (\f{\pi}{a},\f{\pi}{a})$, $\br = a(j_x,j_y)$
and the hole doping rate $\d$,
$$
1-\d(i)=\sum_{\s}\av{\cisd\cis}.
$$
Considering a uniform doping and a uniform staggered magnetization, i.e.,
$\d(i) = \d$ and $m(i) = m$ respectively,
we obtain from (1) the following mean field Hamiltonian for interacting electrons in the square lattice,
\beq
H=-\sumijs \tij \cisd \cjs
+\sumis\left[\f{U}{2}(1-\d)-\f{\s mU}{2}e^{i\bQ\cdot\br}-\m\right]\cisd\cis
+\sumi U\left[(\f{m}{2})^2-\f{1}{4}(1-\d)^2\right].
\eeq
Here the exchange correlation is properly introduced [7].
The third term represents a constant energy
shift.
Ignoring the third term, we rewrite (2) explicitly 
\beqa
H&=&-t\summns\left[\cmpnsd\cmns+\cmmnsd\cmns +e^{-\phc}\cmnpsd\cmns\right.\nonumber
\\
 &&\left.+e^{\phc}\cmnmsd\cmns\right] \nonumber \\
&&-\summns\f{\s mU}{2}e^{i\bQ\cdot\br}\cmnsd\cmns
+\summns\left[\f{U}{2}(1-\d)-\m\right]\cmnsd\cmns
\label{hofs_ham}
\eeqa

Omitting the third term which represents the shift of constant energy,
the Hamiltonian~(\ref{hofs_ham}) is reduced to
\beqa
H&=&-t\summns\left[\cmpnsd\cmns+\cmmnsd\cmns\right.\\\nonumber
   &&\left.+e^{-\phc}\cmnpsd\cmns
+e^{\phc}\cmnmsd\cmns\right]\\\nonumber
   &&-\summns\f{\s mU}{2}e^{i\bQ\cdot\br}\cmnsd\cmns
\eeqa
Now we denote 
$|(j_x, j_y)\ran $ as a single (one) particle state in number representation,
\beq
\ketmn = |1_{(j_x, j_y)}\ran
 = C^\dag_{(j_x, j_y)}|0\ran
\label{single_state}
\eeq
and define
\beqa
|(j_x\!\pm\! 1, j_y)\ran &=&  |1_{(j_x\pm 1, j_y)}\ran 
= C^\dag_{(j_x \pm 1, j_y)} C_{(j_x, j_y)}|1_{(j_x, j_y)}\ran .\nonumber\\
|(j_x, j_y\!\pm\! 1)\ran &=&  |1_{(j_x, j_y\pm 1)}\ran 
= C^\dag_{(j_x, j_y\pm 1)} C_{(j_x, j_y)}|1_{(j_x, j_y)}\ran .
\label{one_ket}
\eeqa
We now use the Schr\"{o}dinger equation, $H|\psi \ran = E|\psi \ran$ 
where $H$ is the Hamiltonian (5) and $\ketpsi$, the eigenstate in number 
representation. We then obtain from the use of (\ref{single_state}) and 
(\ref{one_ket}),
\beqa
&&-t\left[\mmnpsi+\mpnpsi
+e^{-\phc}\mnmpsi+e^{\phc}\mnppsi\right]\\\nonumber
   &&-(-1)^{j_x\!+\!j_y}\f{ mU}{2}\mnpsi=E\mnpsi
\label{hamil}
\eeqa
Here $j_x=1, \cdots , q$ acts as a coordinate in the magnetic cell.
The one-dimensional quasi-periodic boundary condition is satisfied; 
$\lan (j_x, j_y)|\psi\ran =e^{iqk_xa}\langle (j_x\!+\!q, j_y)| \psi\ran$.
There exists
$q$ plaquettes per magnetic unit cell.
Accordingly the energy spectrum has $q$ sub-bands as a result of the applied 
magnetic field.
Here for the range of $k$ vector one magnetic Brillouin zone is equivalent to 
one original Brillouin zone divided by $q$.

The above Hamiltonian (5) is
invariant under the translation $j_y\rarw j_y+2$
for the system of antiferromagnetically correlated electrons.
Thus using the
Bloch theorem,
\beqn
\mnpsi&=&e^{ik_ya j_y} \mzpsi~~~~~~ \mbox{for even $j_y$,} \\
\mbox{and}&&\\
\mnpsi&=&e^{ik_ya j_y} \mopsi~~~~~~ \mbox{for odd $j_y$,}
\eeqn
we write
\beqa
\mnpsi&=&e^{ik_ya j_y}\left(\f{1+(-1)^{j_y}}{2}\mzpsi
       + \f{1-(-1)^{j_y}}{2}\mopsi\right) \\\nonumber
      &=&e^{ik_ya j_y}\left(\f{\mzpsi+\mopsi}{2}
         + (-1)^{j_y}\f{\mzpsi-\mopsi}{2}\right)
\eeqa
By defining 
$$u(j_x)=\f{\mzpsi+\mopsi}{2}, \eqno (10.a)$$
and
$$v(j_x)=(-1)^{-j_x}\f{\mzpsi-\mopsi}{2}, \eqno (10.b)$$
we rewrite (9) above,
\setcounter{equation}{10}
\beq
\mnpsi = e^{ik_ya j_y} \left(u(j_x)+(-1)^{j_x+j_y}v(j_x)\right).
\label{peri}
\eeq

Using (11), we readily find from (8),
\beqa
&& -t \left[u(j_x\!-\!1)+u(j_x\!+\!1)
+\{e^{-\phn}+e^{\phn}\}u(j_x)\right] -\f{ mU}{2}v(j_x)\\\nonumber
&&+(-1)^{j_x+j_y}~t\left[v(j_x\!-\!1)+v(j_x\!+\!1)
+\{e^{-\phn}+e^{\phn}\}v(j_x)\right]\\\nonumber
&&-(-1)^{j_x+j_y}\f{ mU}{2}u(j_x)=E(\left(u(j_x)+(-1)^{j_x+j_y}v(j_x)\right)\nonumber
\eeqa
The expression (12) above leads to the following new Harper's equations which are simply the coupled one-dimensional quasiperiodic difference
equations.
\beqa
&&-t \left[u(j_x\!-\!1)+u(j_x\!+\!1)+2\cos(2\pi \f{p}{q}j_x\! +\!k_ya)u(j_x)\right]-\f{ mU}{2}
v(j_x)=Eu(j_x)\\\nonumber
&&~~~t \left[v(j_x\!-\!1)+v(j_x\!+\!1)+2\cos(2\pi \f{p}{q}j_x\! +\!k_ya)v(j_x)\right]-\f{ mU}{2}
u(j_x)=Ev(j_x)
\eeqa
The coupled difference equations here arise due to the antiferromagnetically correlated electrons.

Defining the reduced energy $\e = E/t$, and the reduced energy gap $\D = mU/t$, 
we rewrite (13) above to explicitly show the following recursion relations,
\beqn
u(j\!+\!1)&=&(-\e-2\cos(2\pi \f{p}{q}j+k_ya))u(j)-u(j\!-\!1)-\f{\D}{2} v(j)\\
v(j\!+\!1)&=&(\e-2\cos(2\pi \f{p}{q}j+k_ya))v(j)-v(j\!-\!1)+\f{\D}{2} u(j)
\eeqn
or in matrix form,
\beqa
\hspace*{-0.5cm}\bmat{c}u(j\!+\!1)\\u(j)\\v(j\!+\!1)\\v(j)\emat
=\bmat{cccc}-\e-2\cos(2\pi \f{p}{q}j\!+\!k_ya)&-1&\f{\D}{2}&0\\
1&0&0&0\\
-\f{\D}{2}&0&\e-2\cos(2\pi \f{p}{q}j\!+\!k_ya)&-1\\
0&0&1&0\emat
\bmat{c}u(j)\\u(j\!-\!1)\\v(j)\\v(j\!-\!1)\emat
\label{matf}
\eeqa
Since there is no possible confusion below, we omit the subscript $x$ of 
$j_x$ in the expression~(\ref{matf}).

To block-diagonalize the square matrix in (\ref{matf}) above,
we introduce the following transformation,
$$
\bmat{c}u'(j)\\v'(j)\emat
=\bmat{cc}\a&\b\\ -\b&\a\emat
\bmat{c}u(j)\\v(j)\emat
$$
where
\beqn
\a&=&\sqrt{\f{1+\sqrt{1-(\f{\D}{2\e})^2}}{2}}\\
\b&=&\sqrt{\f{1-\sqrt{1-(\f{\D}{2\e})^2}}{2}}
\eeqn
Thus the square matrix above is block-diagonalized as
\beq
\bmat{cccc}
-\e_0-2\cos(2\pi \f{p}{q}j\!+\!\n)&-1&0&0\\
1&0&0&0\\
0&0&\e_0-2\cos(2\pi \f{p}{q}j\!+\!\n)&-1\\
0&0&1&0
\emat
\eeq
where $\e_0 \equiv \sqrt{\e^2-(\f{\D}{2})^2}$.

Using (14) and (15) above, we readily find 
$$\bmat{c}u'(j\!+\!1)\\u'(j)\emat =
\bmat{cc}-\e_0-2\cos(2\pi \f{p}{q}j\!+\!k_ya)&-1\\
1&0\emat \bmat{c}u'(j)\\u'(j\!-\!1)\emat \eqno (16.a)$$
$$\bmat{c}v'(j\!+\!1)\\v'(j)\emat =
\bmat{cc}\e_0-2\cos(2\pi \f{p}{q}j\!+\!k_ya)&-1\\
1&0\emat \bmat{c}v'(j)\\v'(j\!-\!1)\emat \eqno (16.b)$$
$\e_0$ here represents the energy spectrum of non-interacting electrons.
The energy dispersion relation \cite{Hong96} for the system of interacting 
electrons is readily obtained from (16) above,
\setcounter{equation}{16}
\beq
 \e(\bk) = \f{U}{2t}(1-\d)-\f{\m}{t}\pm\sqrt{\e_0(\bk)^2 
+ \left(\f{\D}{2}\right)^2},
\label{hofs_corr}
\eeq
with the energy gap, $\f{mU}{2}$. Here we reintroduced the first term which was
omitted in (5).
As seen in (\ref{hofs_corr}) above, the energy spectrum $\e$ shows the variation 
of energy gap with both the staggered magnetization $m$ and the Coulomb 
interaction $U$. 
However the dispersion relation (17) indicates that the self-similarity and 
homeomorphism to the Cantor set are not affected by such electron correlations.

Now using in (16) the translation matrix 
\beq
Q(\e;k_y)\equiv\prod_{j=1}^{q}A_j(\e;k_y),
\label{matq}
\eeq
made of $q$ successive products 
of
$$ A_j(\e;k_y)\equiv 
\bmat{cc} 
-\e_0-2\cos(2\pi \f{p}{q}j\! +\!k_ya)&-1\\
1                              &0 
\emat ,
$$
we obtain
\beq
\bmat{c}g(q\!+\!1)\\g(q)\emat = Q(\e;k_ya)\bmat{c}g(1)\\g(0)\emat 
\eeq
where $g(j) = u'(j)$ or $v'(j)$.
Note that in the expression (\ref{matq}), $\e_0$ is replaced by $\e$
due to relation (\ref{hofs_corr}).
Further we note from the Bloch theorem,
\beq
\bmat{c}g(q\!+\!1)\\g(q)\emat
= e^{iqk_xa } \bmat{c}g(1)\\g(0)\emat .
\eeq 
Thus we readily find from (19) and (20),
$$ \mbox{det}(Q(\e;k_y)-e^{iqk_xa }I) = 0. $$
One of the eigenvalues of $Q$ is complex conjugate to the other in order to 
have a real value of $\tr Q$. Thus we have 
\beq 
\tr Q(\e;k_y) = 2\cos(qk_xa).
\label{trqen}
\eeq
The $k_y$ dependency of the trace 
is additively separable
as shown by Buther and Brown [9], 
and thus we write
\beq
\tr Q(\e)\equiv\tr Q(\e;\f{\pi}{2q} ) = 2\cos(qk_xa) + 2\cos(qk_ya).
\eeq

Since $\tr Q(\e)$ is simply the $q$-th polynomial of $\e$, 
there exist $q$ sub-bands for a given $k$.
Here $k$ is readily obtained for given $\e$.
$\e$ should satisfy the following condition,
\beq
|\tr Q(\e)|\leq 4.
\label{tr_cond}
\eeq
One can determine from the condition (23) whether a state of a given energy $\e$ is possible 
or not.
Using the relations (22) and (23) above, we compute the Hofstadter energy 
spectrum as a function of $\f{p}{q}$ for the square lattice of antiferromagnetically
correlated electrons at half-filling, that is, $\delta = 0$.
In principle, the self consistent equations for $m$ and $\m$ can be derived for a lattice of infinite size [8].
For the case of finite size lattices we obtain $m$ and $\m$ 
numerically  for arbitrary values of $p$ and $q$. 
It is important to thoroughly check numerical precision of the computed 
staggered magnetization $m$ in order to accurately evaluate dispersion 
relation (17).
This is because $m$ in the dispersion relation is very sensitive to the variation of $p/q$.
In order to check numerical accuracy, in Fig.~1 we show the computed results of staggered magnetization 
$m$ as a function of Coulomb repulsion $U$ for the two cases of finite 
($10\times 10$, $20\times 20$ and $30\times 30$ square lattices) and infinite 
size lattices. For both zero magnetic field
and the non-zero magnetic field (corresponding to half a flux quantum
per plaquette),
the computed results of $m$ with the finite size 
lattices (denoted as solid or open circles in Fig.~1) are nearly the same as 
the exact values obtained from the use of the infinite size lattice (denoted 
as a solid line in Fig.~1).
Encouraged by such numerical accuracy, we computed the Hofstadter energy 
spectra of interacting electrons for various values of $U$ by using the
finite size lattice of $10 \times 10$. They are shown in Fig.~2 
through 4.
We find that the structure of the
Hofstadter energy spectrum for interacting electrons is greatly modified only in the  
band gaps and the band widths (the bars in the figures stand for the band width) which
depend on the strength of the Coulomb interaction $U$, as well displayed
 in Figs.~2 through 4.
Indeed this is clearly understood from the expression of the energy dispersion 
relation (17).
Although not shown here, at higher values of $U$ we find no other changes but 
larger band gaps and narrower band widths in the Hofstadter energy spectrum.
The self-similarity of the Hofstadter butterfly is preserved at all values of 
Coulomb interaction $U$. 

The band gap in the structure of the Hofstadter spectrum is predicted to be 
undulatory particularly at the low value of $U=2t$ as shown in Fig.~3.
For the sake fo clarity, undulatory feature of the band gap as a function of 
magnetic flux is explicitly shown in Fig.~5.
It is noted that this undulatory behavior seen at a low value of $U \simeq 2t$
originates from the oscillatory variation of the staggered magnetization $m$ as a function of $p/q$. 
At larger values of $U$ such oscillatory behavior disappears as seen in Fig.~4.
This is because the staggered magnetization (or antiferromagnetic order) becomes increasingly stable and thus insensitive to the variation of magnetic field.
The Hofstadter spectrum yields an additional sub-band only for odd $q$. 
This is caused by the Coulomb interaction which splits the band containing the fermi level 
into two bands.
In short, one can readily see from the dispersion relation (17)
that only the band gaps and the band widths in the Hofstadter spectrum but not 
the essential characters of the self-similarity 
and the Cantor set can be affected by interactions between electrons.

\section{Conclusion}

In the present study we derived a new generalized Harper's equation 
to deal with both the undoped (half-filling, $\d = 0$) and doped ($\d \neq 0$) 
systems by allowing interactions between electrons, and examined the Hofstadter energy 
spectrum as a function of Coulomb repulsion $U$ for the square lattice of correlated 
electrons at half-filling.
We found that such Coulomb interactions between electrons affect only the band 
gaps and band widths in the structure of the Hofstadter butterfly.
Judging from the dispersion relation (17), the inclusion of electron 
correlation effects beyond the mean field level will modify only the band gaps 
and band widths in the Hofstadter spectrum, but not the self-similarity of the 
Hofstadter spectrum and the homeomorphism to the Cantor set.

{\bf Acknowledgement:}

One of us (SHSS) greatly acknowledges the supports of Korea Ministry of 
Education (BSRI 96), Pohang University of Science and Technology (BSRI 96), 
and Center for Molecular Sciences at Korea Advanced Institute of Science and 
Technology. We are also grateful to Seung-Pyo Hong for his assistance.

\newpage
{\Large\bf  Figure Captions}

\begin{itemize}
\item[Fig. 1]Staggered Magnetization with and without magnetic field;
solid line for the infinite size lattice
and dots, for the finite size lattices.

\item[Fig. 2]Hofstadter spectrum of non-interacting electrons at half-filling.

\item[Fig. 3]Hofstadter spectrum of interacting
electrons at half-filling with $U=2t$. 

\item[Fig. 4]Hofstadter spectrum of interacting
electrons at half-filling with $U=4t$.
\item[Fig. 5]Band gap  $\Delta$ as a function of $\f{p}{q}$.
\end{itemize}
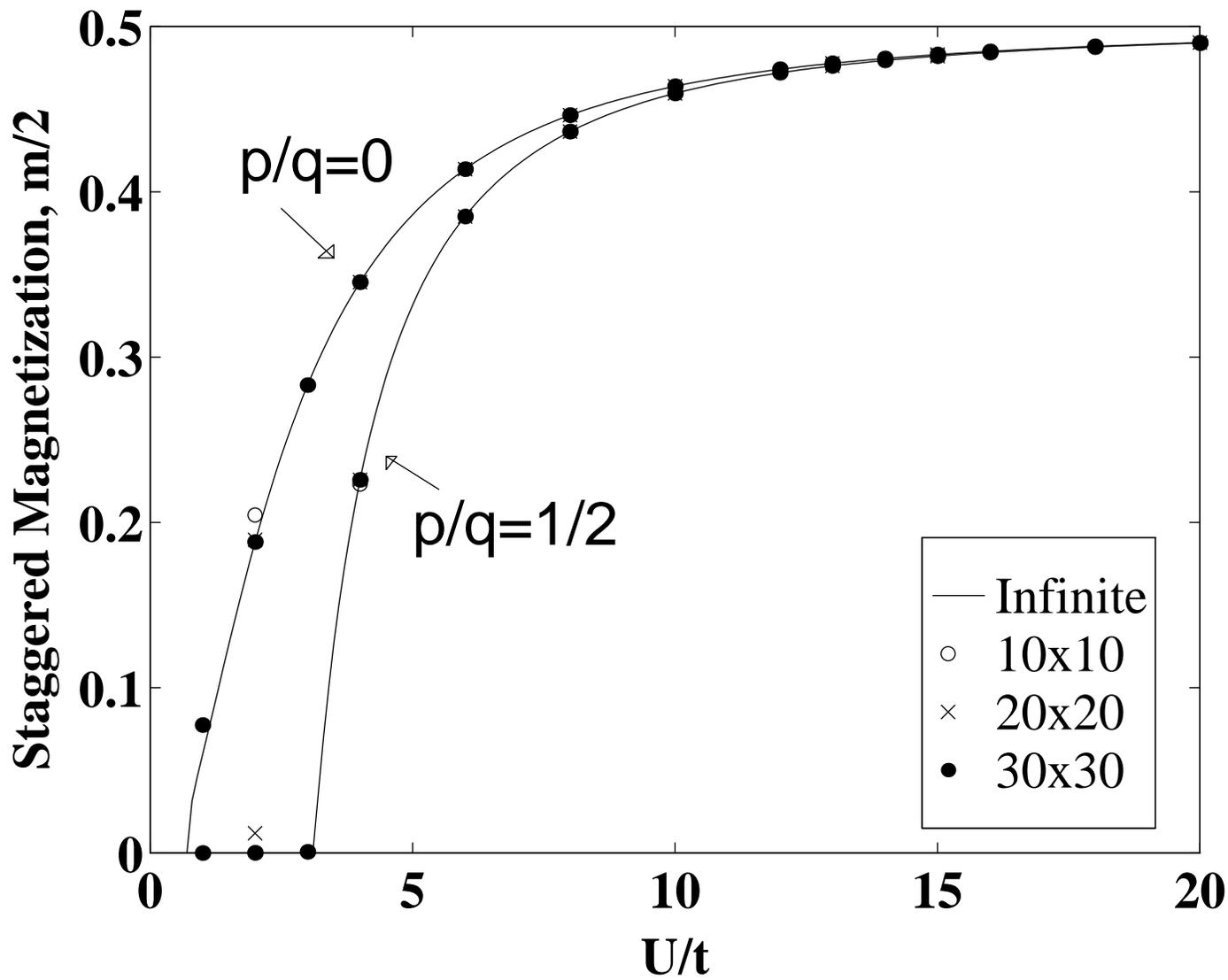
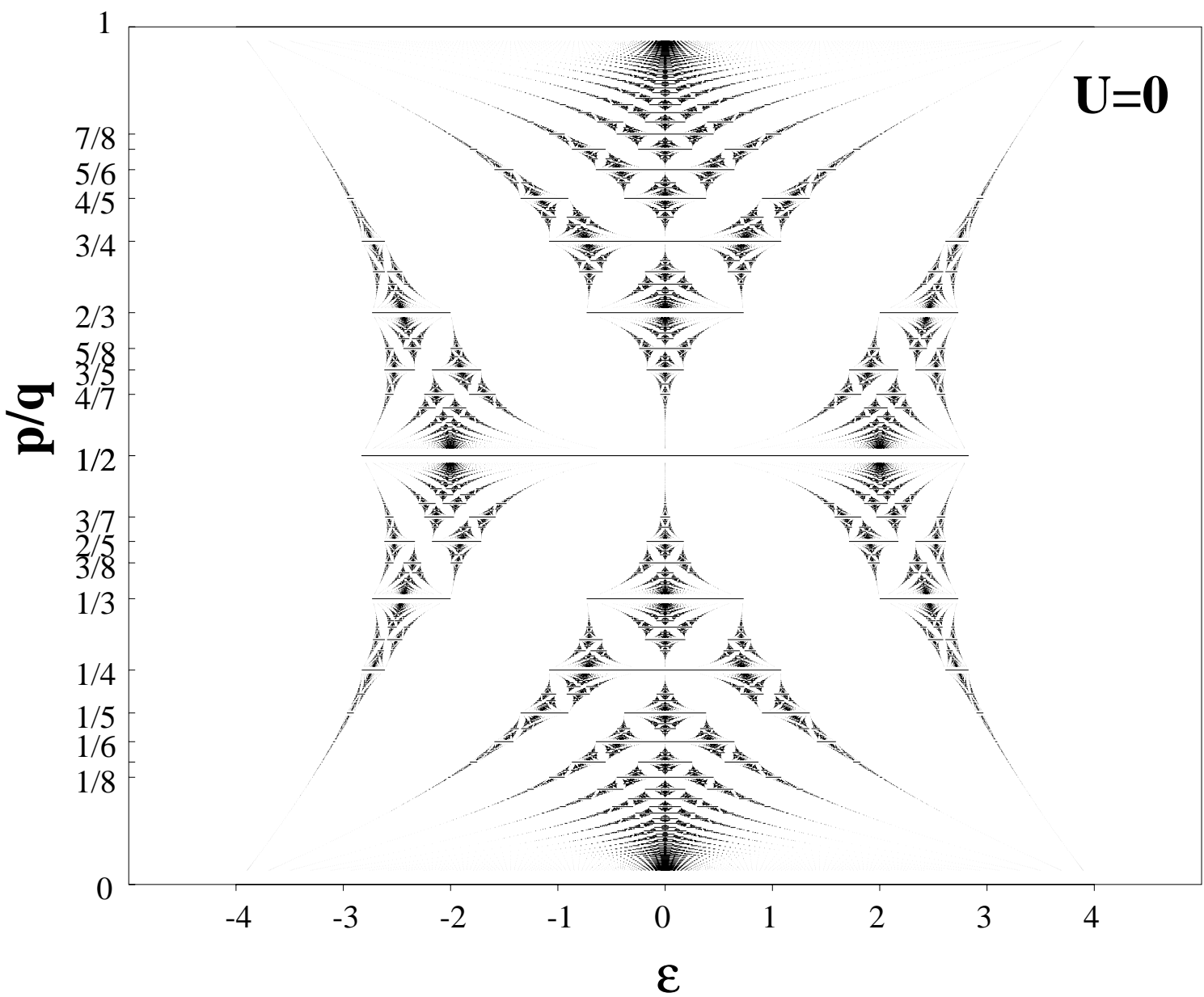
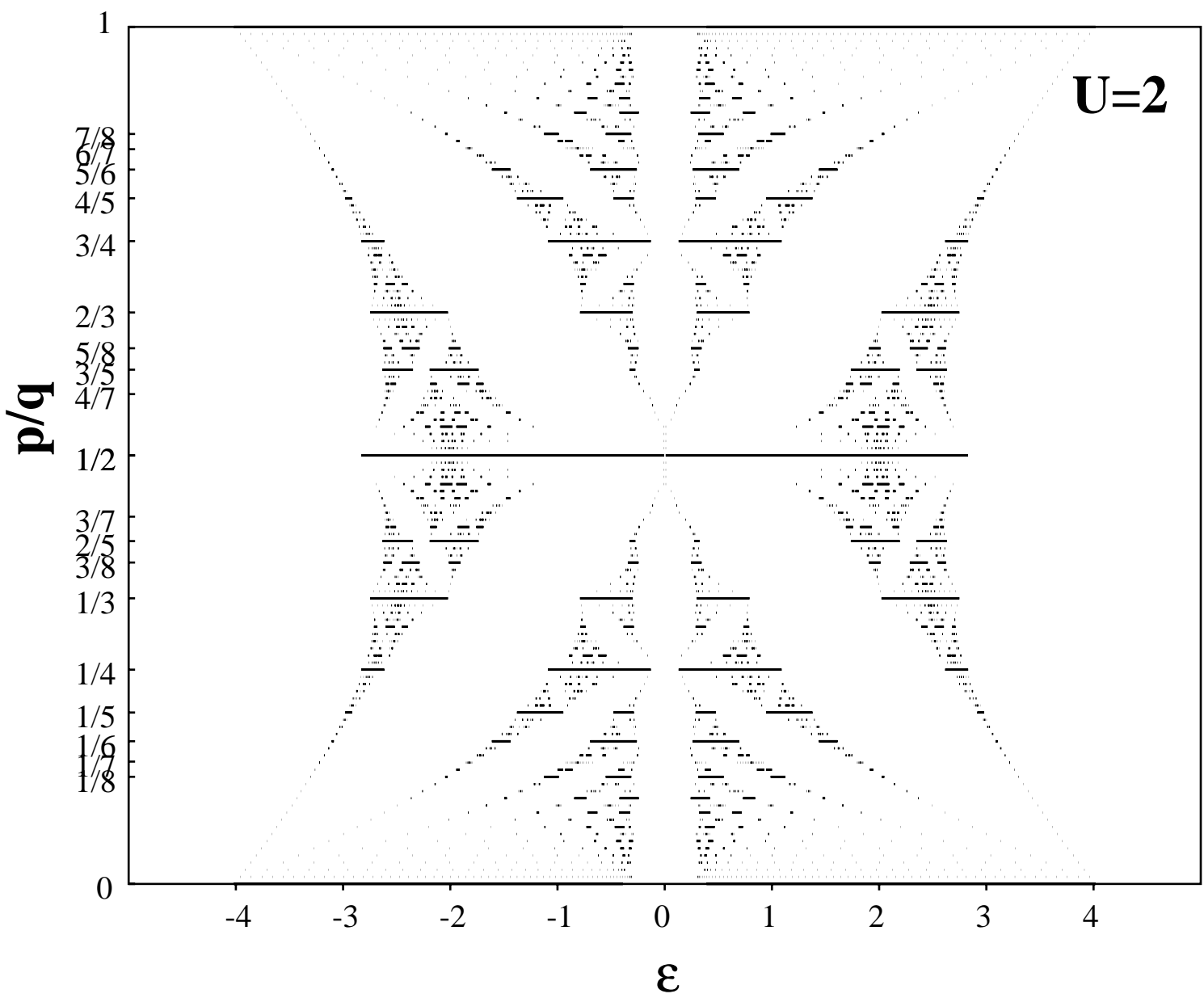
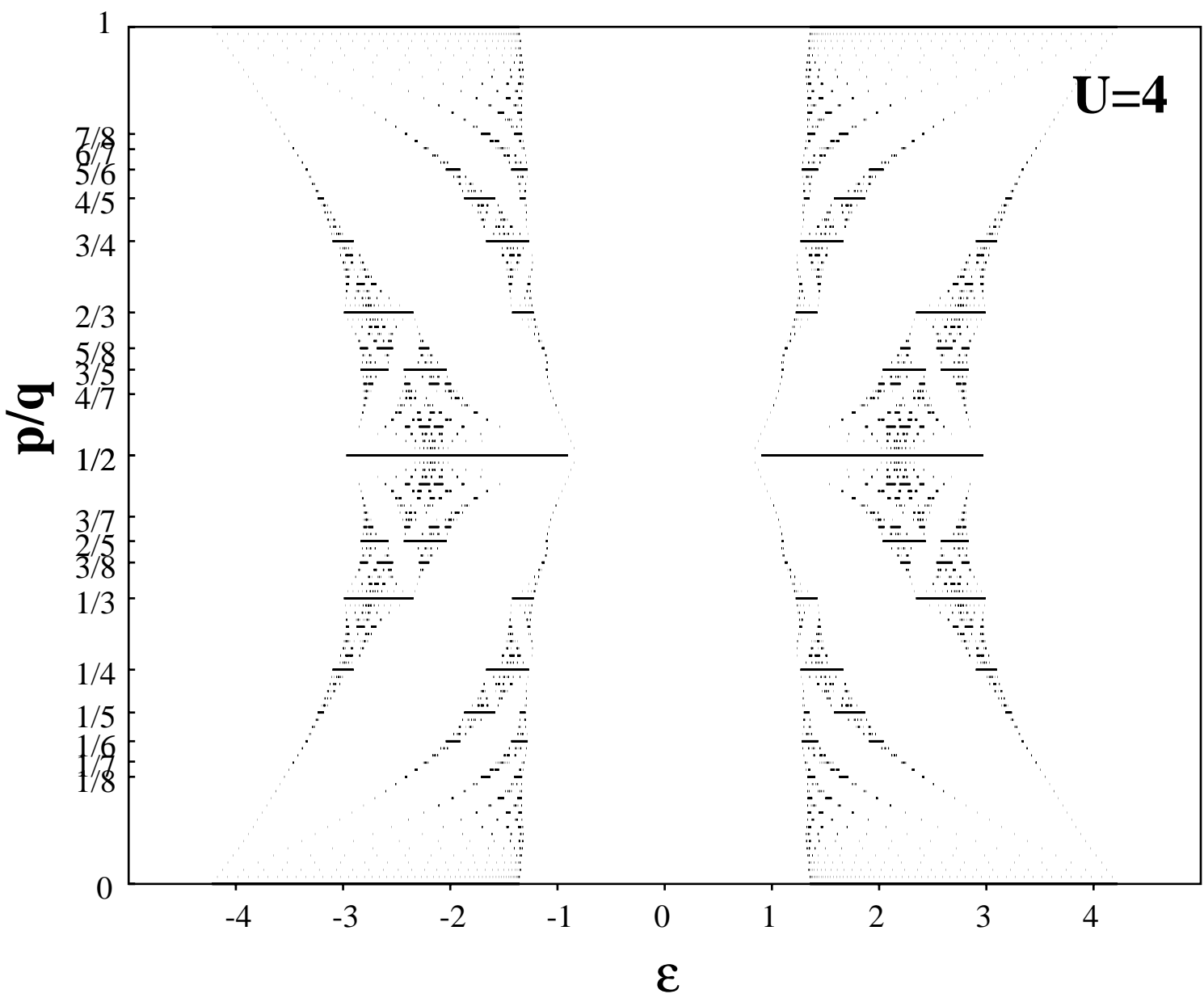
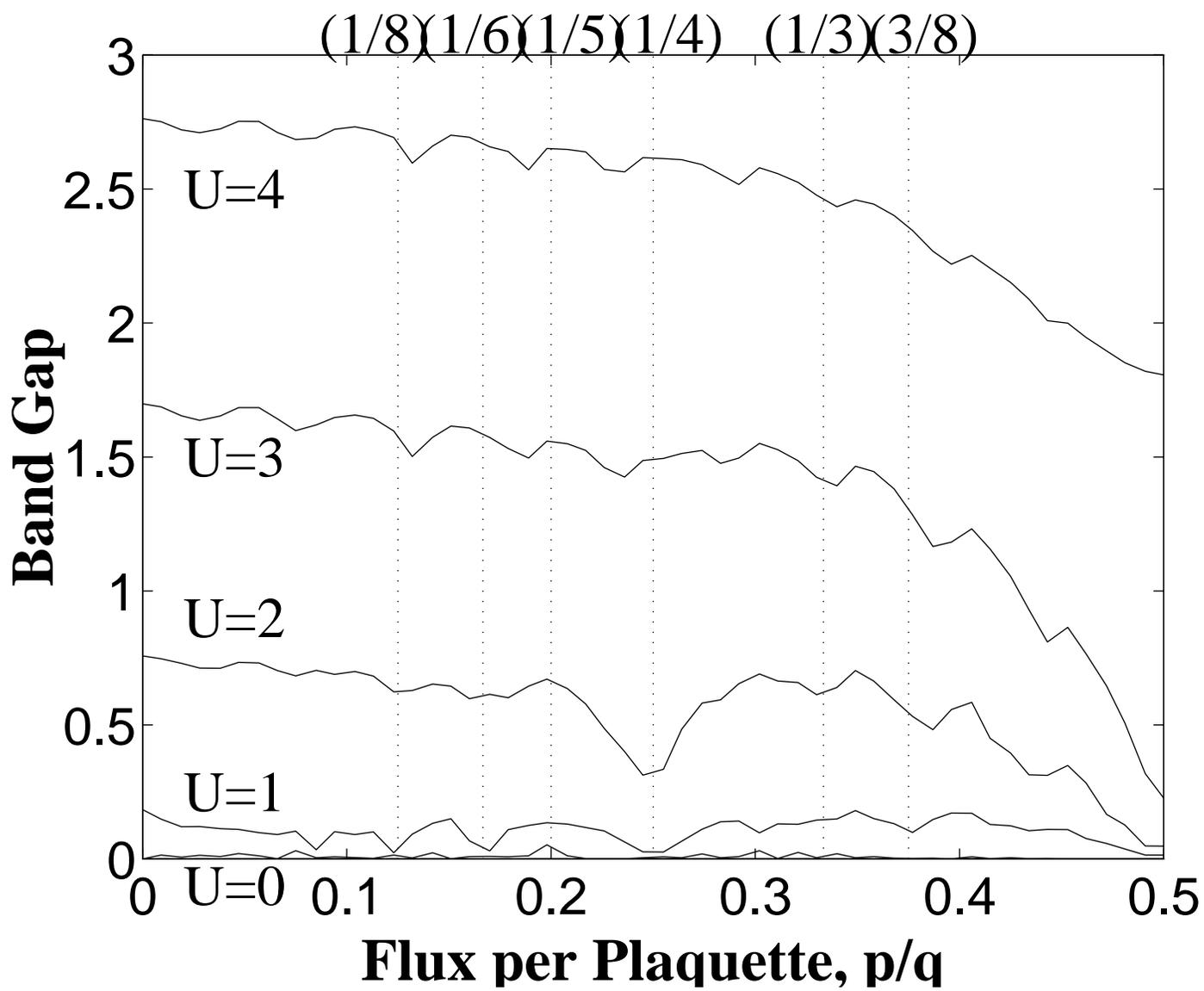
\end{document}